\documentclass[10pt]{article}
\usepackage[titletoc,toc,title]{appendix}
\usepackage{tikz}
\usepackage{graphicx,xcolor,amssymb}
\usepackage{amsmath}

\begin{document}
\title{Lower and upper bounds for configurations of points on a sphere}
\author{Paolo Amore \\
\small Facultad de Ciencias, CUICBAS, Universidad de Colima,\\
\small Bernal D\'{i}az del Castillo 340, Colima, Colima, Mexico \\
\small paolo@ucol.mx 
\and
Ricardo Saenz \\
\small Facultad de Ciencias, CUICBAS, Universidad de Colima,\\
\small Bernal D\'{i}az del Castillo 340, Colima, Colima, Mexico \\
\small rasaenz@ucol.mx}

\maketitle
\begin{abstract}
We present a new proof (based on spectral decomposition) of a bound originally proved by Sidelnikov~\cite{Sidelnikov74},  for the frame potentials 
$\sum_{ij} \left( {\bf P}_i \cdot {\bf P}_j \right)^\ell $ on a unit--sphere in $d$ dimensions.  Sidelnikov's bound is a special case of the lower bound for  the weighted sums $\sum_{ij} f_i f_j \left( {\bf P}_i \cdot {\bf P}_j \right)^\ell$, where  $f_i>0$ are scalar quantities associated to each point on the sphere, which we also prove using spectral decomposition. Moreover, in three dimensions, again using spectral decomposition,  we find a sharp upper bound  for 
$\sum_{ijk}^N \left[ \left( {\bf P}_i  \times {\bf P}_j\right) \cdot {\bf P}_k \right]^2$.

We explore two applications of these bounds: first, we examine configurations of points corresponding to the local minima of the Thomson problem for $N=972$; second, we analyze various distributions of points within a three-dimensional volume, where a suitable weighted sum is defined to satisfy a specific bound.
\end{abstract}

\maketitle

\section{Introduction}
\label{intro}

The study of classical physical many--body systems of interacting particles is extremely challenging, even from a purely numerical point of view. For example, finding the global minimum of the energy of $N$ charges on a unit sphere (a.k.a. 'Thomson problem') becomes rapidly prohibitive when the system is large enough: it is well--known that in this case the number of local minima grows exponentially~\cite{Erber91,Erber97,Calef15, Mehta16}, with a comparable increase in the amount of numerical work required to obtain good candidates for the global  minimum (the best known candidates for global minima  of the Thomson problem over a wide range of $N$ can be found at the Cambridge database~\cite{Walesrepo1, Walesrepo2}). A similar situation occurs when more general interactions (for example the Riesz potential  $V_{ij} = 1/r_{ij}^s$, with $s>0$) are considered (actually ref.~\cite{Calef15} studies the cases  corresponding to $s=0,1,2,3$, where $s=0$ corresponds to a logarithmic potential): in particular the case corresponding to a configuration of $N$ particles on a sphere, with a pairwise interaction $V_{ij} = (\lambda/r_{ij})^s$, and $\lambda>0$ and  $s \rightarrow \infty$, amounts to the  so called 'Tammes problem', i.e. packing on the sphere~\cite{Tammes30}. 

Voronoi diagrams are a useful tool in the classification of  discrete configurations of points on a sphere in three dimensions: Euler's theorem of topology in this case can be cast in terms of the topological charges 
on the sphere (defined as $6-\sigma$, $\sigma$ being the number of sides of the Voronoi cell) and it requires that the total topological charge on the sphere is precisely $12$. As a corollary of this result, it is impossible to
cover the sphere uniquely with hexagons (which carry no topological charge) and at least $12$ pentagons are needed. The nature of the topological defects can change as $N$ grows giving rise to more complex structures
~\cite{Moore97a,Moore97b,Moore99,Wales06,Wales09,Wales13}, but always strictly obeying Euler's theorem.  In general, for low energy configurations of $N$ points on a sphere the number of defects
is much smaller than the total number of particles ($N_{\rm defect} \ll N$).

In the limit $N \rightarrow \infty$ these systems can  be modeled in terms of a continuum elastic medium~\cite{Bowick02,Bowick07,Bowick13,Grason14,Brojan15,Grason16,Grason16b,Grason19,Travesset19,Giomi20,Hilgenfeldt21}. 
Similarly, asymptotic formulas for the energy of configurations of $N$ charges interacting via the Riesz potential for $N \rightarrow \infty$ have been obtained by Saff and collaborators~\cite{Saff94,Saff94b,Saff97,Saff19}. 
Clearly the leading contribution to the energy for $N \rightarrow \infty$ must correspond to a uniform continuum distribution of charge on the sphere, while the subdominant contributions must be sensitive both to the disposition and shape of defects on the surface.  On these grounds one expects a gradual approach to uniformity of the system as $N$ grows (the situation is different in presence of a border, such as for the case a flat disk discussed in \cite{Bedanov94, Bedanov95, Oymak01, Worley06,Moore07,Amore23}).

The main purpose of our paper is to prove general (lower and upper) bounds for configurations of points on a sphere in $d$ dimensions and subsequently  use them to characterize the configurations, quantifying the approach to uniformity.  

In particular,  the bound
\begin{equation}
\begin{split}
\sum_{ij} \left( {\bf P}_i \cdot {\bf P}_j \right)^\ell \geq \frac{\Gamma \left(\frac{d}{2}\right) \Gamma \left(\frac{\ell+1}{2}\right)}{\sqrt{\pi } \Gamma\left(\frac{d+\ell}{2}\right)}  \ \eta_\ell  \ N^2
\label{eq_bound_L}
\end{split}
\end{equation}
holds ($\eta_\ell = 1$ or $0$ if $\ell$  is either even or odd) for $N$ points, ${\bf P}_i$ ($i=1,\dots,N$), arbitrarily distributed on a unit sphere in $d$ dimensions. 
This bound has been originally proved long time ago by Sidelnikov~\cite{Sidelnikov74} (see also ref.~\cite{Waldron18}) , but we provide here an alternative proof
based on spectral decomposition.

Using a similar approach we have been able to generalize the bound to the case of weighted sums $\sum_{ij}  f_i f_j \ \left( {\bf P}_i \cdot {\bf P}_j \right)^\ell $ ($f_i>0$, $i=1,\dots,N$):
\begin{equation}
\begin{split}
\sum_{ij}  f_i f_j \ \left( {\bf P}_i \cdot {\bf P}_j \right)^\ell \geq  \frac{\Gamma \left(\frac{d}{2}\right) \Gamma \left(\frac{\ell+1}{2}\right)}{\sqrt{\pi } \Gamma\left(\frac{d+l}{2}\right)}  \ \eta_\ell  \ \left(\sum_{i=1}^N f_i  \right)^2 
\label{eq_bound_L_generalized}
\end{split}
\end{equation}
and to derive the upper bound (we have only studied the case in three dimensions)
\begin{equation} 
\begin{split}
\sum_{ijk}^N \left[ \left( {\bf P}_i  \times {\bf P}_j\right) \cdot {\bf P}_k \right]^2  \leq \frac{2}{9} N^3  \ .
\label{eq_bound_U}
\end{split}
\end{equation}
To the best of our knowledge both bounds, eq.~(\ref{eq_bound_L_generalized}) and eq.~(\ref{eq_bound_U}) are new.

In particular notice that the bound in (\ref{eq_bound_L_generalized}) can be applied to configurations of points in the interior of  a sphere in $d$ dimensions, by relating  the weights $f_i$  to the distances of the points from the center of the sphere.  Closeness to satisfy the bounds may signal a larger degree of uniformity of the distribution of points.  A possible application of this tool could be in the study of hyperuniform systems~\cite{Torquato03,Torquato16,Torquato18}.

The paper is organized as follows: in section \ref{bound_sid} we prove the bounds of eqs.~(\ref{eq_bound_L}) and (\ref{eq_bound_L_generalized}) using spectral decomposition (a lower bound to the Riesz energy is then obtained applying Sidelnikov's bounds); in section \ref{sec_upper_bounds} we use the spectral approach  to prove a new upper bound for points on the sphere in $d=3$ dimensions, eq.~(\ref{eq_bound_U}); in section \ref{sec_applications} we apply our bounds to study low energy configurations  in the Thomson problem for $N=972$, as well as distributions of points in two and three dimensions.  Finally, in section \ref{sec_conclusion} we draw our conclusions.

\section{Lower bounds}
\label{bound_sid}

We consider configurations of $N$ points arbitrarily distributed on a unit sphere in $d$ dimensions, ${\bf P}_i$, with $i=1,\dots,N$ and
define 
\begin{equation}
\hat{\bf FP}_\ell \equiv
\sum_{i=1}^N \sum_{j=1}^{N}  \left({\bf P}_i \cdot {\bf P}_j \right)^{\ell}  \hspace{0.5cm} , \hspace{0.5cm} \ell =1,2,\dots  \  ,
\label{eq_frame_potential}
\end{equation}
which is known as  the $\ell$-frame potential~\cite{Benedetto03}.

For a continuous distribution the expression above corresponds to 
\begin{equation}
\begin{split}
\widehat{\bf \mathcal{FP}}_\ell &\equiv \int \int    \rho(\Omega_d) \rho(\Omega_d')  \left({\bf P} \cdot {\bf P}' \right)^{\ell} d\Omega_d d\Omega_d' 
\hspace{0.5cm} , \hspace{0.5cm} \ell =1,2,\dots  \ , 
\end{split}
\label{eq_mom_cont_d}
\end{equation}
where $\rho(\Omega_d)$  is a density with the proper normalization ($\int \rho(\Omega_d) d\Omega_d = N$) and $d\Omega_d$ is the solid angle element in $d$ dimensions. 
In what follows we will use the notation $\widetilde{\bf \mathcal{FP}}_\ell $  for the special case where $\rho(\Omega_d)$ is constant (uniform distribution).

We will first consider the case of the unit circle, where the position of a point is determined by a single angle. We start by considering  
a continuous distribution on the circle with arbitrary density
\begin{equation}
\rho(\theta) = c_0^{(1)} \psi_0^{(1)}(\theta)+ \sum_{k=1}^\infty \sum_{a=1}^2 c_k^{(a)} \psi^{(a)}_k(\theta) \ ,
\end{equation}
where 
\begin{equation}
\psi_k^{(a)}(\theta) = \left\{ \begin{array}{ccc}
1/\sqrt{2\pi} & , & k = 0 ,  a = 1 \\
\cos(k \theta)/\sqrt{\pi} & , & k > 0 , a = 1 \\
\sin(k \theta)/\sqrt{\pi} & , & k > 0 , a = 2 \\
\end{array}\right.
\end{equation}
is the orthonormal basis with periodic boundary conditions. The choice $c_0^{(1)} = N/\sqrt{2\pi}$ ensures the correct normalization of the density. 
For $k=0$ there is only the element corresponding to $a=1$: to make things simpler (having to distinguish between the cases $k=0$ and $k \neq 0$) 
we can set $c_0^{(2)} = 0$  and $\psi_0^{(2)} = 0$, avoiding unnecessary complications.
 
The identities
\begin{equation}
\cos^\ell x = \frac{1}{2^{\ell}} \sum_{j=0}^k\binom{\ell}{j} \cos ((\ell-2 j) x) 
\end{equation}
and
\begin{equation}
\pi \sum_{a=1}^2 \psi_k^{(a)}(\theta) \psi_k^{(a)}(\phi) = \cos ( k (\theta-\phi))  \hspace{1cm} , \hspace{1cm} k=1,2,\dots
\end{equation}
can be used  to write
\begin{equation}
\begin{split}
\cos^\ell (\theta-\phi)  &= \frac{\pi}{2^{\ell-1}} \sum_{j=0}^{\left\lfloor \frac{\ell}{2}\right\rfloor} \sum_{a=1}^2 \binom{\ell}{j} \psi_{\ell-2j}^{(a)}(\theta) \psi_{\ell-2j}^{(a)}(\phi) 
\end{split} \ .
\end{equation}

In this way we obtain
\begin{equation}
\begin{split}
\widehat{\bf \mathcal{FP}}_\ell  &=  \sum_{j=0}^{\left\lfloor \frac{\ell}{2}\right\rfloor}  \sum_{a=1}^2  \frac{\pi}{2^{\ell-1}} \binom{\ell}{j} \left[ c_{\ell-2 j}^{(a)}\right]^2  \ .
\label{eq_fpl_cont}
\end{split}
\end{equation}

By looking at this expression we see that 
\begin{equation}
\begin{split}
\widehat{\bf \mathcal{FP}}_\ell &= \left\{ 
\begin{array}{ccc}
2^{-\ell} \binom{\ell}{\frac{\ell}{2}} N^2 + \frac{\pi}{2^{\ell-1}} \binom{\ell}{\frac{\ell}{2}-1} (c_{2,1}^2 +c_{2,2}^2 ) +\dots & , & \ell \ \ {\rm even} \\
\frac{\pi}{2^{\ell-1}} \binom{\ell}{\frac{\ell-1}{2}} (c_{1,1}^2 +c_{1,2}^2 ) +\dots & , & \ell \ \ {\rm odd} \\
\end{array}
\right. \ .
\end{split}
\end{equation}

The case of a uniform distribution corresponds to set all the coefficients $c_{k,a}=0$ with $k>0$ in the formula above:  from (\ref{eq_fpl_cont}) follows 
\begin{equation}
\widehat{\bf \mathcal{FP}}_\ell \geq \widetilde{\bf \mathcal{FP}}_\ell  = 2^{-\ell} \binom{\ell}{\frac{\ell}{2}} N^2 \ .
\end{equation}

Now consider an arbitrary distribution of $N$ points on the circle: by using the completeness of the basis we can express the density  as
\begin{equation}
\rho(\theta) = \sum_{p=1}^N \delta(\theta-\theta_p) = \sum_{p=1}^N \left[ \psi_{0}^{(1)}(\theta) \psi_{0}^{(1)}(\theta_p)  +
\sum_{k=1}^\infty \sum_{a=1}^2 \psi_{k}^{(a)}(\theta) \psi_{k}^{(a)}(\theta_p) \right] \ ,
\end{equation}
and one can read off the coefficients $c_{k}^{(a)}$ of the expansion directly from this expression
\begin{equation}
c_{k}^{(a)} = \sum_{p=1}^N  \psi_{k}^{(a)}(\theta_p) \nonumber \ .
\end{equation}

For a uniform distribution of points on the unit circle, $\theta_p  = \frac{2\pi p}{N}$, the property
\begin{equation}
\sum_{p=1}^N \psi_k^{(a)}(\frac{2\pi p}{N}) = \frac{N}{\sqrt{2\pi}} \delta_{k,0}\delta_{a,1}
\end{equation}
implies that the corresponding density is also uniform, $\rho(\theta) = \sum_{p=1}^N \delta(\theta -\frac{2\pi p}{N}) = \frac{N}{2\pi}$, and therefore
the bound is saturated, hence
\begin{equation}
	\widehat{\bf \mathcal{FP}}_\ell = \widetilde{\bf \mathcal{FP}}_\ell  \nonumber \  .
\end{equation}

These results can be extended to three dimensions. In this case we consider an arbitrary continuous distribution of charge over the sphere with a density  $\rho(\Omega)$:
\begin{equation}
\rho(\Omega) = \sum_{l=0}^\infty \sum_{m=-l}^l C_{lm} Y_{lm}(\Omega) \ ,
\end{equation}
where $Y_{lm}(\Omega)$ are the spherical harmonics. The coefficient $C_{00}$ in this expansion is obtained from the normalization, $C_{00} = N/\sqrt{4 \pi}$; additionally 
as $\rho$ is real valued, he have $C_{l,-m} = (-1)^m C_{l,m}^\star$.

We can use the addition theorem for spherical harmonics
\begin{equation}
P_l (cos\gamma) = \frac{4\pi}{2l +1} \sum_{m=-l}^{l} Y_{l m }(\Omega) Y^\star_{l m }(\Omega') \ ,
\end{equation} 
where $\cos\gamma \equiv {\bf P} \cdot {\bf P}'$ and the property 
\begin{equation}
\begin{split}
\cos^l \gamma 
&= \sum_{k=0}^l \Gamma_{\ell, k} \sum_{j=-k}^{k} (-1)^j Y^\star_{k -j }(\Omega) Y^\star_{k j}(\Omega') \\
\end{split}
\end{equation}
with $\Gamma_{\ell,k} \equiv 4\pi  l! \ \eta_{l-k} /\left( 2^{(l-k)/2} \left( (l-k)/2 \right)! (l+k+1)!!\right)$, to write
\begin{equation}
\begin{split}
\widehat{\bf \mathcal{FP}}_\ell &\equiv \int \int    \rho(\Omega_d) \rho(\Omega_d')  \cos^\ell\gamma  d\Omega_d d\Omega_d'  
=  \sum_{k=0}^\ell \sum_{j=-k}^{k}  \Gamma_{\ell,k} \ \ |C_{k,j}|^2 \ .
\end{split}
\end{equation}

This formula implies that
\begin{equation} 
\widehat{\bf \mathcal{FP}}_\ell =  \frac{N^2}{1+ \ell} \eta_\ell + \mathcal{R}_\ell    \ ,
\end{equation}
where $\mathcal{R}_\ell \geq 0$  contains the squared norms of the coefficients that are not related to the normalization. 
Therefore we conclude that the frame potential of a continuous uniform distribution constitutes a lower bound to the frame potential of
arbitrary continuous distributions:
\begin{equation}
\widehat{\bf \mathcal{FP}}_\ell \geq \widetilde{\bf \mathcal{FP}}_\ell =  \frac{N^2}{1+ \ell} \eta_\ell\  .
\end{equation}

As before we can consider a discrete distribution as described by the density
\begin{equation}
\rho(\Omega) = \sum_{j=1}^N \delta(\Omega -\Omega_j) = \sum_{j=1}^N \sum_{l=0}^\infty \sum_{m=-l}^l Y_{lm}(\Omega) Y^\star_{lm}(\Omega_j) \ ,
\end{equation}
where the coefficients $C_{l,m}$ can be easily read off. We conclude that $ \widetilde{\bf \mathcal{FP}}_\ell$ provides a lower bound also to arbitrary discrete distributions.  

The generalization to higher dimensions can be carried in an analogous way, by considering the spherical harmonic decomposition of $L^2(\mathbb S^{d-1})$,
where $\mathbb S^{d-1}$ is the $(d-1)$-dimensional sphere in $\mathbb R^d$.  Indeed, one has $L^2(\mathbb S^{d-1}) = \bigoplus_{l\ge 0} H_l$, where $H_l$
is the space of spherical harmonics of degree $l$ on the unit sphere, defined as the restrictions of harmonic $l$-degree polynomials on $\mathbb R^d$ to  $\mathbb S^{d-1}$.

Each space $H_l$ has dimension $d_l = (2l+d-2)\dfrac{(l+d-3)!}{l!(d-2)!}$ and, if $Y_{l,1}, Y_{l,2}, \ldots, Y_{l,d_l}$ is an orthonormal basis for $H_l$, then 
\begin{equation}\label{relsGeg}
\sum_{m=1}^{d_l}  Y_{l,m}^\star(x)Y_{l,m}(y) = F_l(x\cdot y),
\end{equation}
where $F_l$ is a computable polynomial of degree $l$ in one variable (see \cite{Folland20}). In fact,
\[
F_l(t) = \frac{2l+d-2}{(d-2)\omega_d}C_l^{(d-2)/2}(t),
\]
 with $\omega_d$ the surface area of the $(d-1)$-sphere in $\mathbb{R}^d$. $C_l^\nu(t)$ in this expression are the Gegenbauer polynomials
\[
C_l^\nu(t) = \sum_{k=0}^{\lfloor l/2 \rfloor}
(-1)^k 2^{l-2k} \frac{\Gamma(\nu + l - k)}{\Gamma(\nu)k!(l-2k)!} t^{l-2k}.
\]
(see \cite{Lebedev72}). 

One can recursively solve equations \eqref{relsGeg} to obtain an expression of
the form
\begin{equation}\label{power}
(x\cdot y)^l = \sum_{k=0}^{\lfloor l/2\rfloor} B_k^l
\sum_{m=1}^{d_{l-2k}} Y_{l-2k,m}^\star(x) Y_{l-2k,m}(y),
\end{equation}
with $B_k^l\ge 0$, which are analogous to to equation (22).

Again, consider an arbitrary density distribution over $\mathbb S^{d-1}$
\[
\rho(\Omega) = \sum_{l=0}^\infty \sum_{m=1}^{d_l} c_{l,m} Y_{l,m}(\Omega) \ ,
\]
where $c_{0,0} = N/\omega_d$,  with $\omega_d$ the surface area of the $(d-1)$-sphere in $\mathbb{R}^d$.

Then
\begin{equation}
\label{bound_d}
\begin{split}
\widehat{\mathcal{FP}}_l  &=
\int\int ({\bf P} \cdot {\bf P}')^l \rho(\Omega) \rho(\Omega') d\Omega d\Omega'\\
&= \sum_{k=0}^{\lfloor l/2 \rfloor} \sum_{j=1}^{d_{l-2k}} |c_{l-2k,j}|^2 B_k^l
\ge \widetilde{\mathcal{FP}}_l,
\end{split}
\end{equation}
where $\widetilde{\mathcal{FP}}_l$ is the frame potential with respect to the
uniform density $\rho\equiv N/\omega_d$. For odd $l$, $\widetilde{\mathcal{FP}}_l = 0$,
and for even $l$, we have $\widetilde{\mathcal{FP}}_l = B_{l/2}^l \ N^2$, where
the $B_{l/2}^l$ can be calculated recursively by the equations
\[
\sum_{k=0}^{l/2}b_{l/2-k}^l B_k^{2k} = 0,
\]
where the $b_k^l$ are the coefficients of the polynomial $F_l$, i.e.
\[
F_l(t) = \sum_{k=0}^{l/2} b_k^l y^{l-2k}.
\]
One can verify that 
\begin{equation}\label{genFP}
B_{l/2}^l = \frac{(l-1)!!(d-2)!!}{(l+d-2)!!} = \frac{\Gamma \left(\frac{d}{2}\right) \Gamma \left(\frac{\ell+1}{2}\right)}{\sqrt{\pi } \Gamma\left(\frac{d+l}{2}\right)}  , 
\end{equation}
where $m!!$ is de double factorial of $m$, $m!! = m(m-2)(m-4)\ldots$. 

Eq.~(\ref{bound_d}) with the coefficients specified in (\ref{genFP}) provides a lower bound to the $\ell$-frame potential for an arbitrary continuous 
distribution on the sphere in $d$ dimensions. This bound can be extended to the discrete case, by  exploiting the completeness of the basis, as
previously done in two and three dimensions. By doing this one recovers the formula obtained by Sidelnikov in \cite{Sidelnikov74} (see also \cite{Waldron18}).

The definition of the $\ell$--frame potentials of eq.~(\ref{eq_frame_potential} ) can be extended to the case where at each point on the sphere is associated a scalar or vector quantity:
\begin{equation}
\hat{\bf FP}^{(f)}_\ell \equiv \sum_{i=1}^N \sum_{j=1}^{N}  f_i f_j \ \left({\bf P}_i \cdot {\bf P}_j \right)^{\ell}  \hspace{0.5cm} , \hspace{0.5cm} \ell =1,2,\dots  
\label{eq_frame_potential_extended}
\end{equation}
and
\begin{equation}
\hat{\mathbf{FP}}^{({\bf f})}_\ell  \equiv \sum_{i=1}^N \sum_{j=1}^{N}  \left( {\bf f}_i \cdot {\bf f}_j \right) \ \left({\bf P}_i \cdot {\bf P}_j \right)^{\ell}  \hspace{0.5cm} , \hspace{0.5cm} \ell =1,2,\dots  \  ,
\label{eq_frame_potential_vector_extended}
\end{equation}
respectively~\footnote{Notice that the scalar case for $f_i = 1$ correspond to the original definition.}. These expressions follow from using the scalar and  vector densities 
\begin{equation}
\begin{split}
\rho_f(\Omega) &= \sum_{i=1}^N f_i \delta(\Omega-\Omega_i)  \\
\vec{\rho}_f(\Omega) &= \sum_{i=1}^N {\bf f}_i \delta(\Omega-\Omega_i)  \\
\end{split}
\end{equation}
and obey the normalization conditions
\begin{equation}
\begin{split}
\int \rho_f(\Omega)  d\Omega &= \sum_{i=1}^N f_i  \\
\int \vec{\rho}_f(\Omega)  d\Omega &= \sum_{i=1}^N {\bf f}_i  \\
\end{split} \ .
\end{equation}

It is straightforward to derive the corresponding bounds for the generalized frame potentials defined above: for example, for $d=3$, one has
\begin{equation}
\hat{\bf FP}^{({\bf f})}_\ell = \sum_{k=0}^\ell \sum_{j=-k}^k  \Gamma_{\ell, k} \  {\bf c}_{k,j}^\star \cdot \ {\bf c}_{k,j}
\end{equation}
where ${\bf c}_{k,j} = \sum_{i=1}^N {\bf f}_i  \ Y_{k,j}^\star(\Omega_i)$ and
\begin{equation}
{\bf c}_{0,0}^\star \cdot {\bf c}_{0,0} = \sum_{a=1}^N  \sum_{b=1}^N \frac{{\bf f}_a \cdot {\bf f}_b}{4\pi} \ .
\end{equation}

Morover, since ${\bf c}_{k,j}^\star \cdot \ {\bf c}_{k,j} \geq 0$ for all values of $k$ and $j$, we obtain the bound
\begin{equation}
\begin{split}
\sum_{ij}  {\bf f}_i\cdot {\bf f}_j \ \left( {\bf P}_i \cdot {\bf P}_j \right)^\ell \geq \frac{\eta_\ell}{\ell+1}   \
\left(\sum_{i,j=1}^N {\bf f}_i\cdot {\bf f}_j  \right) \ .
\label{eq_bound_L_generalized_vec_3}
\end{split}
\end{equation}

This result can be generalized to arbitrary integer values of $d$ as done before, thus yielding the bound
\begin{equation}
\begin{split}
\sum_{ij}  {\bf f}_i\cdot {\bf f}_j \ \left( {\bf P}_i \cdot {\bf P}_j \right)^\ell \geq \frac{\Gamma \left(\frac{d}{2}\right) \Gamma \left(\frac{\ell+1}{2}\right)}{\sqrt{\pi } \Gamma\left(\frac{d+l}{2}\right)}  \ \eta_\ell  \
\left(\sum_{i,j=1}^N {\bf f}_i\cdot {\bf f}_j  \right) \ .
\label{eq_bound_L_generalized_vec}
\end{split}
\end{equation}

An analogous calculation yields eq.~(\ref{eq_bound_L_generalized}) for the simpler scalar case.

We briefly discuss now how these lower bounds  can be used to derive a lower bound for the Riesz-energy:
consider a system of $N$ point charges on a unit sphere in $d=3$ dimensions, interacting via a Riesz potential
\begin{equation}
V_{ij} =  \frac{\text{sign}(s) }{r_{ij}^s} \hspace{1cm} , \hspace{1cm} s \neq 0 \ , \ s \geq -1 \ ,
\end{equation}
where $r_{ij}$ is the euclidean distance between two charges
\begin{equation}
r_{ij} = \sqrt{2 (1 - {\bf P}_i \cdot {\bf P}_j)} \ .
\end{equation}

The total energy  of a system of charges interacting with the Riesz potential is
\begin{equation}
	E(s) = \frac{1}{2} \sum_{i ,j=1}^N   \frac{\text{sign}(s) }{r_{ij}^s} \ .
	\end{equation}

For the case $s=-1$, we express the total energy of the system as
\begin{equation}
E(-1) = -\frac{1}{2} \sum_{i ,j=1}^N r_{ij} = - \frac{1}{\sqrt{2}}  \sum_{k=0}^\infty
\left(\begin{array}{c}
1/2\\
k\\
\end{array} \right) (-1)^k  \sum_{i,j=1}^N ({\bf P}_i\cdot {\bf P}_j)^k \ ,
\end{equation}
where we take advantage of the property $r_{ii}=0$.  Because the frame potentials appearing in this expression
obey the lower bounds  that we have previously derived we have the bound over the total energy of the system
\begin{equation}
E(-1) \geq - \frac{2}{3} N^2 \ ,
\end{equation}
which corresponds to the leading asymptotic (in $N$) behavior of the energy found in eq.(4.6) of ref.~\cite{Saff94b}~\footnote{The different sign  comes from the fact that 
\cite{Saff94b} deals with a maximization problem, whereas we consider a minimization problem.}. 

The derivation of the bounds for the total energy is more delicate for  $s \geq 0$ and it requires a modification of the inter particle distance:
\begin{equation}
r_{ij}(\delta) \equiv \sqrt{2 \left (1  + \delta - {\bf P}_i \cdot {\bf P}_j\right)} \hspace{1cm} ,  \hspace{1cm} \delta \geq 0 
\end{equation}
and of the inter--particle potential as
\begin{equation}
V_{ij}(\delta) = \frac{1}{r_{ij}^s(\delta)} \hspace{1cm} , \hspace{1cm} s > 0 \ .
\end{equation}

We discuss explicitly the special case $s=1$ and write the total energy of a system of $N$ points as
\begin{equation}
\widehat{\mathbf{E}}(1,\delta) = \frac{1}{2} \sum_{i \neq j} \frac{1}{r_{ij}(\delta)} =  \frac{1}{2} \left[ \sum_{i,j} \frac{1}{r_{ij}(\delta)} - 
\frac{N}{ (2\delta )^{1/2}} \right] \  ,
\label{eq_energy_discrete}
\end{equation}
 where the finiteness of $r_{ii}(\delta)$ for $\delta>0$ is exploited to include the diagonal terms into the sum.  Because 
$r_{ij}(\delta) \geq r_{ij}(0)$ for $\delta >0$ we have that  $\widehat{\mathbf{E}}(1,\delta) < \widehat{\mathbf{E}}(1,0)$.

The binomial theorem allows us to cast the energy in terms of the frame potentials as
\begin{equation}
\begin{split}
\widehat{\mathbf{E}}(1,\delta)  &= \frac{1}{2\sqrt{2}} \sum_{\ell=0}^\infty  (-1)^\ell  \binom{-\frac{1}{2}}{\ell}  
\frac{\hat{\bf FP}_\ell}{ (\delta +1)^{l+\frac{1}{2}}} -  \frac{N}{ 2 \sqrt{2\delta} }  \ .
\label{eq_energy_discrete2}
\end{split}
\end{equation}

Let $\widehat{\mathcal{E}}(s,\delta)$  be the energy corresponding to (\ref{eq_energy_discrete2}) where the bound (\ref{eq_bound_L}) is saturated; 
a simple calculation  yields
\begin{equation}
\begin{split}
\widehat{\mathcal{E}}(1,\delta) &= \frac{N}{4} \left(\frac{2 N}{\sqrt{\delta +\sqrt{\delta  (\delta+2)}+1}}-\sqrt{\frac{2}{\delta }}\right) \\
\end{split}
\label{eq_lower_bound_energy}
\end{equation}
with
\begin{equation}
\begin{split}
\widehat{\mathbf{E}}(1,0) >  \widehat{\mathbf{E}}(1,\delta) & >  \widehat{\mathcal{E}}(1,\delta) \ .
\end{split}
\end{equation}
 
By calling $\delta^\star$ be the value at which $ \widehat{\mathcal{E}}(1,\delta) $ is maximal:
\begin{equation}
\delta^\star = \frac{4}{4 N-\sqrt{8 N+1}-1} \ ,
\end{equation}
we finally obtain the bound
\begin{equation}
\begin{split}
\widehat{\mathbf{E}}(1,0) >  \widehat{\mathcal{E}}(1,\delta^\star)   \ , 
\end{split}
\end{equation}
where for $N \rightarrow \infty$   
\begin{equation}
\widehat{\mathcal{E}}(1,\delta^\star)  \approx \frac{N^2}{2}-\frac{N^{3/2}}{\sqrt{2}}+\frac{N}{8}+\frac{\sqrt{N}}{16 \sqrt{2}}+\frac{1}{64} +\dots \ .
\end{equation}

The term $N^2/2$ is the electrostatic energy of a uniformly charged unit sphere, as expected; the next term, $\frac{N^{3/2}}{\sqrt{2}}$, agrees with the 
asymptotic behavior of the energy for $N \gg 1$ (see \cite{Saff97}), but the  numerical coefficient exceeds the value found in numerical calculation ~\cite{Erber91, Glasser92,Saff94b} and also conjectured to be $0.55305$ in \cite{Saff97} .

\section{An upper bound}
\label{sec_upper_bounds}

In this section we derive a different  upper bound, which does not rely on (\ref{eq_bound_L}) . 

To this purpose we define the quantity
\begin{equation}
\hat{\bf AFP}_\ell\equiv  \sum_{i,j,k=1}^N \sum_{a,b,c=1}^3  \left( \epsilon_{abc} {\bf P}_{a}(\theta_1)   {\bf P}_{b}(\theta_2) {\bf P}_{c}(\theta_3) \right)^\ell
\end{equation}
where $\epsilon_{a,b,c}$ is the antisymmetric Levi-Civita tensor. We will call ${\bf AFP}_\ell$ the $\ell$- -antisymmetric frame potential.

Similarly we consider its continuous counterpart 
\begin{equation}
\widehat{\bf \mathcal{AFP}}_\ell \equiv   \int \int \int \rho(\Omega_1)  \rho(\Omega_2) \rho(\Omega_3)  \sum_{a,b,c=1}^3  \left( \epsilon_{abc} {\bf P}_{a}(\theta_1)   {\bf P}_{b}(\theta_2) {\bf P}_{c}(\theta_3) \right)^\ell  \ d\Omega_1  d\Omega_2  d\Omega_3\ .
\end{equation}
Notice that $\hat{\bf AFP}_\ell$ trivially vanishes if $\ell$ is odd so that only even values of $\ell$ are interesting.

We first consider the case of a homogeneous distribution, $\rho(\Omega) = N/4\pi$. In this case we extend the notation previously adopted for the frame potentials and indicate with 
$\widetilde{\bf \mathcal{AFP}}_\ell$ the $\ell$--antisymmetric frame potential corresponding to a uniform distribution.

One can calculate explicitly these quantities for different even values of $\ell$:
\begin{equation}
\begin{split}
\widetilde{\bf \mathcal{AFP}}_2 &= \frac{2 N^3}{9}  \\
\widetilde{\bf \mathcal{AFP}}_4 &= \frac{8 N^3}{75} \\
&\dots
\end{split}
\label{eq_AFP_uniform}
\end{equation}

These expressions appear to correspond to the general formula 
\begin{equation}
\widetilde{\bf \mathcal{AFP}}_\ell  = \frac{2 \sqrt{\pi } N^3}{(\ell-1) (\ell+1)^2}  \ \frac{\Gamma \left(\frac{\ell}{2}+1\right)}{ \Gamma \left(\frac{\ell}{2}-\frac{1}{2}\right)} \ .
\label{eq_AFP_uniform_2}
\end{equation}

Let us now discuss the case of a non--uniform continuous distribution, for  $\ell = 2$. To start with we need to decompose the 
integrand in the basis of spherical harmonics as
\begin{equation}
\begin{split}
\left[\left( {\bf P}(\theta) \times {\bf P}(\phi) \right) \cdot {\bf P}(\chi) \right]^2 = \sum_{\left\{ l,m\right\}} \kappa_{l_1,m_1,l_2,m_2,l_3,m_3} Y_{l_1,m_1}(\Omega_\theta) 
 Y_{l_2,m_2}(\Omega_\phi)  Y_{l_3,m_3}(\Omega_\chi)  \ ,
\end{split}
\label{eq_non_uniform_afp}
\end{equation}
where $\sum_{\left\{ l,m\right\}} \equiv  \sum_{l_1=0}^2 \sum_{m_1 =-l_1}^{l_1}  \sum_{l_2=0}^2 \sum_{m_2 =-l_2}^{l_2}  \sum_{l_3=0}^2 \sum_{m_3 =-l_3}^{l_3}$. 
The  explicit form of the coefficients $\kappa$ in this expansion is reported in the appendix \ref{sec:appA}.

By  using the orthonormality of the spherical harmonics we obtain
\begin{equation}
\begin{split}
\hat{\bf \mathcal{AFP}}_2 &=   \frac{2 N^3}{9}-\frac{8}{15} \pi  N \left(C_{2,0}^2+2 |C_{2,1}|^2 +2 |C_{2,2}|^2 \right) \\
&+ \frac{16 \pi ^{3/2}}{45 \sqrt{5}} \left(6 \left(\left|C_{2,1}\right|^2 -2 \left| C_{2,2} \right|^2 \right) C_{2,0}  \right. \\
&+ \left. 3 \sqrt{6} \left(C^\star_{2,2}  C_{2,1}^2+C_{2,2} \left( C^\star_{2,1}\right)^2\right)+2 C_{2,0}^3\right) 
\end{split}  \ . 
\label{eq_FPA_1}
\end{equation}

In the discrete case the coefficients $C_{l,m}$ correspond to
\begin{equation}
\begin{split}
C_{l,m} = \sum_{i=1}^N Y_{l,m}(\Omega_i)  \ .
\end{split}
\end{equation}
It is easy to see that $C_{2,0}$ fulfills the bounds
\begin{equation}
-\frac{N}{4} \sqrt{\frac{5}{\pi }}  \leq C_{2,0} \leq \frac{N}{2} \sqrt{\frac{5}{\pi }}  \ ,
\label{eq_bounds}
\end{equation} 
corresponding respectively to choosing $\theta_i = \pi/2$ (lower bound) and $\theta_i = 0$ (upper bound).

It is convenient to reorganize eq.~(\ref{eq_FPA_1}) as
\begin{equation}
\begin{split}
\hat{\bf AFP}_2 &=   \left[ \frac{2 N^3}{9}-\frac{8 \pi N}{15} C_{2,0}^2+\frac{32 \pi ^{3/2}}{45 \sqrt{5}} C_{2,0}^3 \right] \\
&-\frac{32}{75} \pi ^{3/2} \left| C_{2,1} \right|^2 \left(\frac{5 N}{2 \sqrt{\pi   }}-\sqrt{5} C_{2,0}\right)  \\
&-\frac{32}{75} \pi ^{3/2} \left| C_{2,2} \right|^2 \left(2 \sqrt{5} C_{2,0}+\frac{5 N}{2 \sqrt{\pi }}\right) \\
&+  \frac{16}{5} \sqrt{\frac{2}{15}} \pi ^{3/2} \left(C_{2,1}^2 {C^\star}_{2,2}+C_{2,2} {C^\star}_{2,1}^2\right)
\end{split}
\label{eq_FPA_2}
\end{equation}
and look at these terms separately.

The first term has a local maximum at $C_{2,0}=0$ and a local minimum at $C_{2,0} =\frac{N}{2} \sqrt{\frac{5}{\pi}}$, where it vanishes. 
It also vanishes at $C_{2,0} = -\frac{N}{4} \sqrt{\frac{5}{\pi }}$. 

Therefore
\begin{equation}
0 \leq  \left[ \frac{2 N^3}{9}-\frac{8 \pi N}{15} C_{2,0}^2+\frac{32 \pi ^{3/2}}{45 \sqrt{5}} C_{2,0}^3 \right] \leq  \frac{2}{9} N^3  \ .
\end{equation}

Assuming that $C_{2,0}=0$ we have
\begin{equation}
\begin{split}
\hat{\bf AFP}_2 &=   \frac{2 N^3}{9} -\frac{16}{15} \pi N (\left| C_{2,1} \right|^2  +\left| C_{2,2} \right|^2)  \\
&+  \frac{16}{5} \sqrt{\frac{2}{15}} \pi ^{3/2} \left(C_{2,1}^2 {C^\star}_{2,2}+C_{2,2} {C^\star}_{2,1}^2\right) \ .
\end{split}
\label{eq_FPA_3}
\end{equation}

By expressing $C_{l,m} = |C_{l,m} | e^{i \xi_{l,m}}$ we have
\begin{equation}
\begin{split}
\hat{\bf AFP}_2 &=   \frac{2 N^3}{9} -\frac{16}{15} \pi N (\left| C_{2,1} \right|  - \left| C_{2,2} \right|)^2 \\
&+  \frac{32}{75} \pi |C_{2,1}| \ |C_{2,2}|  \left(-5N + \sqrt{30\pi} |C_{2,1}| \ \cos(2\xi_{21} -\xi_{22})\right) \ .
\end{split}
\label{eq_FPA_4}
\end{equation}

However from the definition of $C_{l,m}$ one observes that  $|C_{2,1}| \leq \frac{N}{4} \sqrt{\frac{15}{2 \pi }}$, which can be used to obtain the bound 
\begin{equation}
\begin{split}
\hat{\bf AFP}_2 &\leq   \frac{2 N^3}{9} -\frac{16}{15} \pi N (\left| C_{2,1} \right|  - \left| C_{2,2} \right|)^2 \\
&+  \frac{32}{75} N \pi |C_{2,1}| \ |C_{2,2}|  \left(-5  + \frac{15}{4}  \ \cos(2\xi_{21} -\xi_{22})\right) \\
&\leq \frac{2 N^3}{9} = \widetilde{\bf \mathcal{AFP}}(2)
\end{split}
\label{eq_FPA_5}
\end{equation}

At this point  the reader probably expects that the proof above can be extended to $\ell = 4, 6, \dots$: a simple check on the Platonic solids shows that for the tetrahedron and the  cube 
$\widetilde{\bf \mathcal{AFP}}_\ell < \hat{\bf AFP}_\ell$ for $\ell= 4,6,8$. For the octahedron the situation is even worse, because $\widetilde{\bf \mathcal{AFP}}_\ell < \hat{\bf AFP}_\ell$
for all even $\ell > 2$.

\section{Applications}
\label{sec_applications}

We now consider two applications of the bounds obtained in this paper: to the Thomson problem and to volume distributions of points in three dimensions.

\begin{figure}
\begin{center}
\includegraphics[width=4cm]{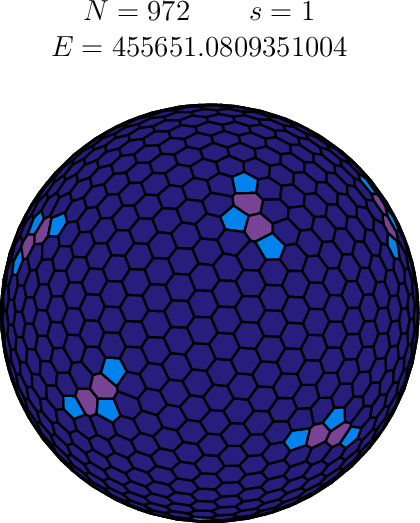}    
\caption{Voronoi diagram for the candidate global minimum configuration for $N=972$}
\label{Fig_voro_972_1}
\end{center}
\end{figure}

As a first application we apply the bounds  to study $1000$ low energy configurations for the Thomson problem with $N=972$.  These configurations corresponds to local minima obtained by perturbing the candidate for global minimum available at the Cambridge database~\cite{Walesrepo1} and subsequently minimizing  the total energy. In Fig.~\ref{Fig_voro_972_1}  we display the  Voronoi diagram for the candidate to global minimum (notice that the color of the  cells depends on the number of sides).

\begin{figure}
\begin{center}
\bigskip
\includegraphics[width=7cm]{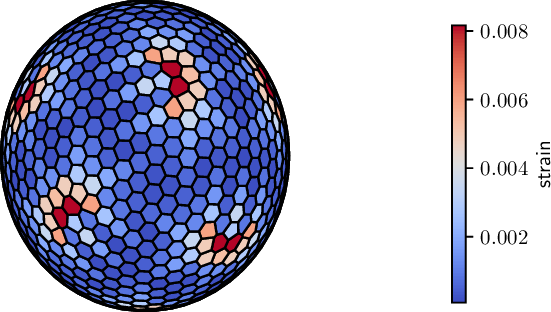}    
\bigskip
\caption{Strain intensity  for the candidate global minimum configuration for $N=972$}
\label{Fig_voro_972_2}
\end{center}
\end{figure}

\begin{figure}
\begin{center}
\bigskip
\includegraphics[width=7cm]{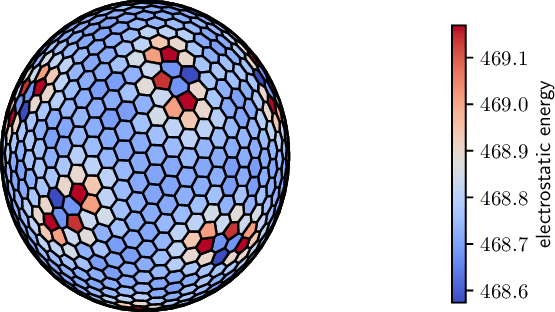}    \\
\bigskip
\caption{Electrostatic energy  for the candidate global minimum configuration for $N=972$}
\label{Fig_voro_972_3}
\end{center}
\end{figure}

\begin{figure}
\begin{center}
\bigskip
\includegraphics[width=10cm]{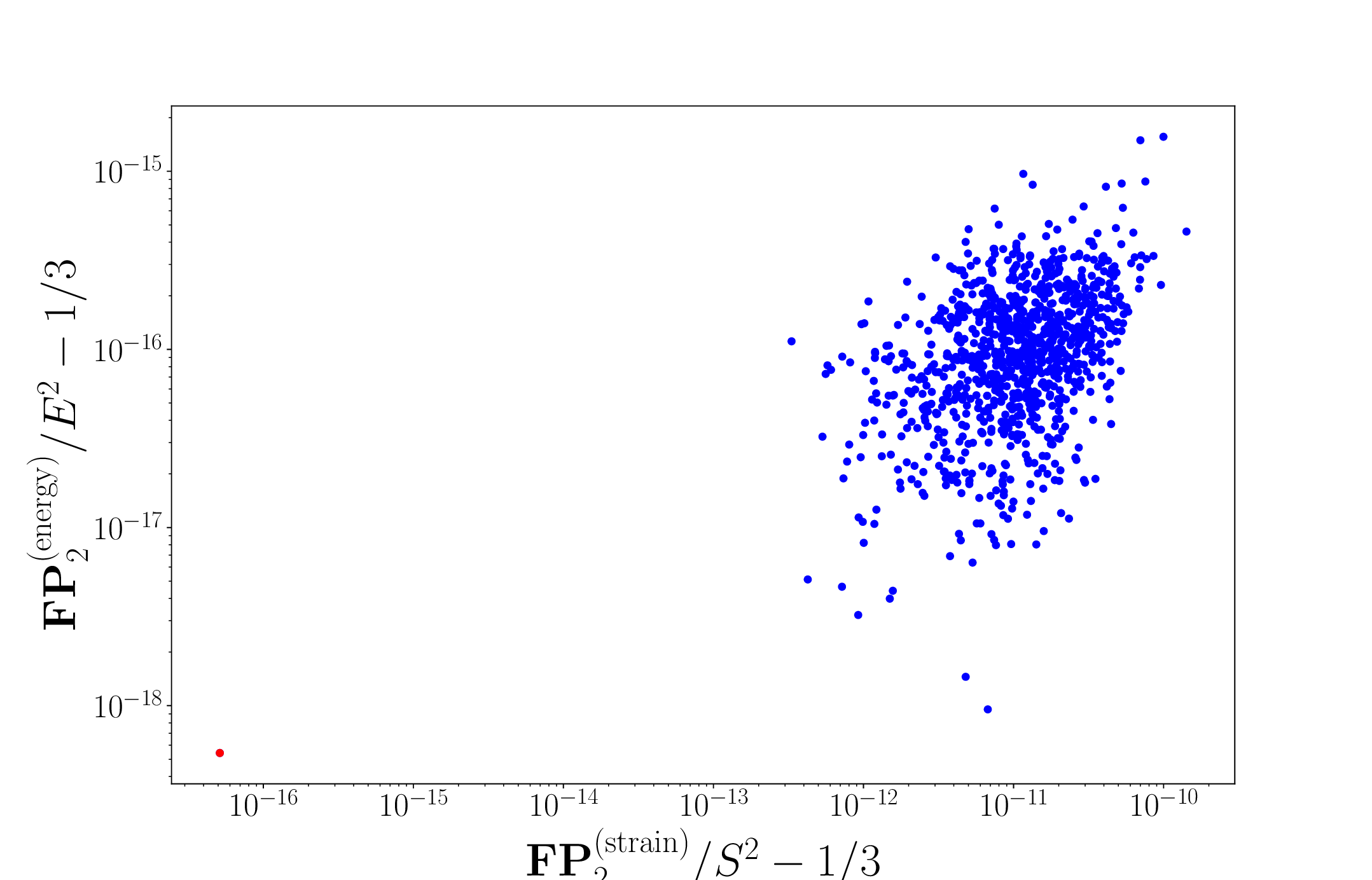} 
\bigskip
\caption{Weighted frame potentials of strain versus energy for the $1000$ configurations obtained slightly perturbing the candidate global minimum (in red). $S$ and $E$ are the total strain and total electrostatic energy of a configuration. }
\label{Fig_972_strain_vs_energy_FP}
\end{center}
\end{figure}

To conduct our numerical experiments we have calculated for all configurations the individual "strain" and electrostatic energy of the Voronoi cells (the strain
is here defined as the distance between the point charge generating the cell and the center of mass, calculated from its vertices --  clearly, for a regular spherical  polygon the strain should vanish). The intensities of the strain and of the electrostatic energy for the global minimum  are displayed in Figs.~\ref{Fig_voro_972_2} and \ref{Fig_voro_972_3}, respectively. One can appreciate that for the lowest energy configuration the strain is mostly concentrated in the region of defects, particularly in the heptagonal cells and to a lesser extent in the cells immediately close. In particular from Fig.~\ref{Fig_972_strain_vs_energy_FP} we see that the lowest energy configuration (red point in the figure) is the closest to saturate the bounds ${\bf FP}_2^{(strain)}=S^2/3 $  and ${\bf FP}_2^{(energy)}=E^2/3$ among the different configurations considered. Because the strain is prevalent in the region where defects are present, we can argue that the defects tend to distribute uniformly on the sphere, to a good degree  (Euler's theorem requires that   the total topological charge on the sphere {\sl must} equal $12$, without providing information on its spatial distribution).

One can more specifically obtain a bound for the topological defects, by considering  the  weighted frame potential $\sum_{i,j}^N f_i f_j \left( \vec{\bf P}_i \cdot \vec{\bf P}_j \right)^\ell$, with $f_i = 0$ for hexagonal Voronoi cells and $f_i = 1$ for all remaining cells.

In this case  the following bound must be obeyed
\begin{equation}
\sum_{i,j}^{N_{\rm defect}} \left( \vec{\bf P}_i \cdot \vec{\bf P}_j \right)^\ell  > \frac{N_{\rm defect}^2}{\ell+1} \eta_\ell \hspace{0.5cm} , \hspace{0.5cm} \ell = 1,2,\dots ,
\end{equation}
where $N_{\rm defect}$ is the total number of non--hexagonal cells .

\begin{figure}
\begin{center}
\bigskip
\includegraphics[width=10cm]{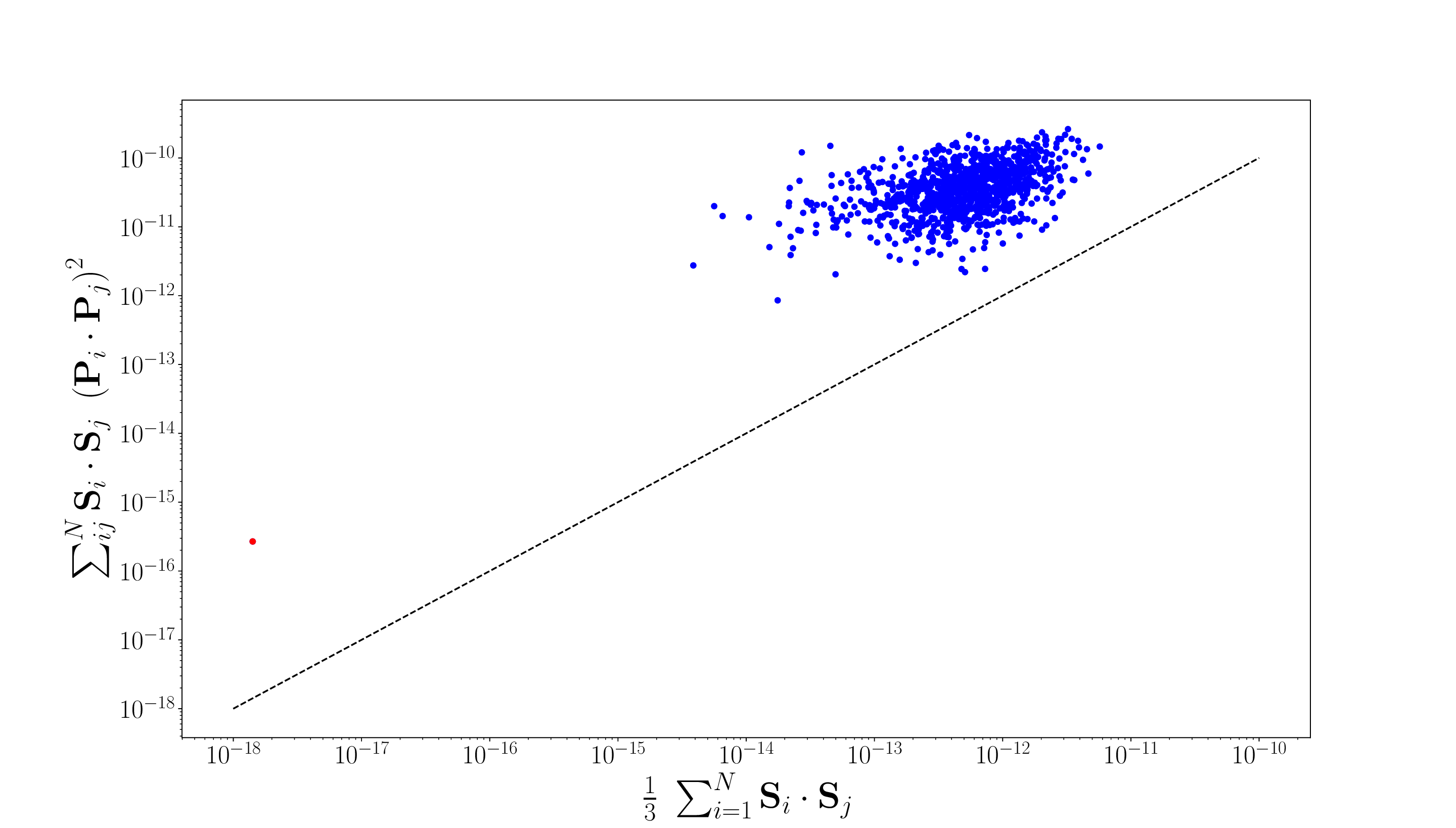} 
\bigskip
\caption{$\left( \sum_{ij}^N {\bf S}_i \cdot {\bf S_j} \right)/3$ vs $\sum_{ij} {\bf S}_i \cdot {\bf S_j}  ({\bf P}_i \cdot {\bf P}_j)^2$ for the $1000$ configurations obtained slightly perturbing the candidate global minimum (in red). ${\bf S}_i$ is  the strain vector at point ${\bf P}_i$. }
\label{Fig_972_strain_FP_vec}
\end{center}
\end{figure}

In Fig.~\ref{Fig_972_strain_FP_vec} we have considered the vector frame potential  $\sum_{ij} {\bf S}_i \cdot {\bf S}_j  \ \left( {\bf P}_i \cdot {\bf P}_j\right)^2$: the dashed line in the figure corresponds to the lower bound $\frac{1}{3} \sum_{ij}^N  {\bf S}_i \cdot {\bf S}_j $, which is almost saturated for all the configurations under consideration.

\newpage

\begin{figure}	
	\begin{center}
		\begin{tikzpicture}[scale=0.4]
			\def\radius{4} 
			\def\squareSize{8} 
			
			\draw[thick] (0,0) circle (\radius);
			
			
			\def\step{1} 
			
			\foreach \x in {-15,...,15} {
				\foreach \y in {-15,...,15} {
					\pgfmathsetmacro{\hx}{\x * \step + \y * \step * 0.5}
					\pgfmathsetmacro{\hy}{\y * \step * 0.866} 
					
					\pgfmathparse{(\hx >= -\squareSize) && (\hx <= \squareSize) && (\hy >= -\squareSize) && (\hy <= \squareSize)}
					\ifnum\pgfmathresult=1
					\pgfmathsetmacro{\distanceSquared}{\hx*\hx + \hy*\hy}
					\pgfmathparse{\distanceSquared <= \radius*\radius}
					\ifnum\pgfmathresult=1
					\fill[blue] (\hx, \hy) circle (0.1);
					\else
					\fill[red] (\hx, \hy) circle (0.1);
					\fi
					\fi
				}
			}
				
			\fill[red] (0, 0) circle (0.1); 
			
		\end{tikzpicture}
		\caption{Hexagonal lattice in two dimensions: the points inside the circle are used to build a weighted sum, for which appropriate bounds are established. The central point is discarded to avoid undefined quantities. }
		\label{Fig_hexa}
	\end{center}
\end{figure}
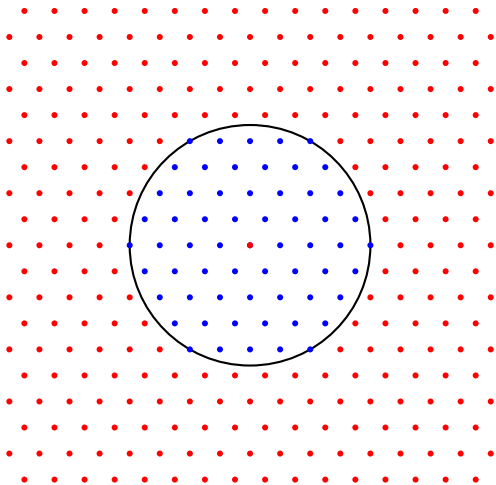

A second application the we consider is to a distribution of points in a volume in two and three dimensions. For simplicity we first consider the two dimensional case, as illustrated in Fig.~\ref{Fig_hexa}: the points of a hexagonal lattice falling inside a circle of radius $r$ are selected (the blue points in the figure) and used to define a weighted sum~\footnote{Any point in the origin must be discarded from the set to avoid undefined quantities.}
\begin{equation}
{\bf FP}_\ell^{(r,\gamma)} \equiv \sum_{i,j=1}^N \left(	\frac{|{\bf P}_i|}{r} \right)^\gamma \ \left(	\frac{|{\bf P}_j|}{r} \right)^\gamma \  
\left( \frac{{\bf P}_i \cdot {\bf P}_j}{|{\bf P}_i| |{\bf P}_j|} \right)^\ell \  ,
\end{equation}
which obeys the bound
\begin{equation}
{\bf FP}_\ell^{(r,\gamma)} \geq  \eta_\ell	2^{-\ell} \left(\begin{array}{c}
\ell \\
\frac{\ell}{2}
\end{array}\right) \left[\sum_{i=1}^N \left(	\frac{|{\bf P}_i|}{r} \right)^\gamma\right]^2 \ , 
\end{equation}
where $\gamma \geq 0$ is an arbitrary parameter that changes the weights in the sum.

We define
\begin{equation}
	\Delta_\ell^{(r,\gamma)} \equiv \frac{1}{N^2} \left[ {\bf FP}_\ell^{(r,\gamma)} - \eta_\ell	2^{-\ell} \left(\begin{array}{c}
		\ell \\
		\frac{\ell}{2}
	\end{array}\right) \left[\sum_{i=1}^N \left(	\frac{|{\bf P}_i|}{r} \right)^\gamma\right]^2 \right] \geq 0 \ .
\end{equation}

For the hexagonal lattice in Fig.~\ref{Fig_hexa}, with the circle centered on a lattice point (which is excluded from the sum), the bound is saturated for odd values $\ell$ and for $\ell= 2, 4$, regardless of the value of $\gamma$. The saturation of the bound is a manifestation of the uniformity of the lattice.  The circle provides an observation window, similarly to the one  used in \cite{Torquato18} (see Fig.1  of that paper) to study the number variance as a function of the size of the window: in the present case, we can obtain useful information both by determining the smallest value of $\Delta_\ell^{(r,\gamma)}$  and studying its  variability as the  observation window of fixed size, is displaced around, provide information on the distribution.

Similarly in  three dimensions $\Delta_\ell^{(r,\gamma)} $ is defined as 
\begin{equation}
\Delta_\ell^{(r,\gamma)} \equiv \frac{1}{N^2} \left[ {\bf FP}_\ell^{(r,\gamma)} -	\frac{1}{\ell+1} \left(\sum_{i=1}^N \left(	\frac{|{\bf P}_i|}{r} \right)^\gamma\right)^2 \right] \geq 0 \ .
\end{equation}

Again, for a cubic lattice, where the sphere is centered at the center of the unit cell, $\Delta_\ell^{(r,\gamma)}$  saturates the bound for $\ell=2$ and arbitrary $r$.

\section{Conclusions}
\label{sec_conclusion}

In  this paper we have obtained a new proof of a lower bound over $\ell$-frame potentials, using spectral decomposition; this bound was  originally found by Sidelnikov~\cite{Sidelnikov74}, but using our approach we have been able to generalize the bound to a larger class of weighted (scalar and vector) $\ell$-frame potentials (which includes the former as a special case). Still using spectral decomposition and introducing
a new family of antisymmetric frame potentials, we have found a special upper bound in $d=3$ dimensions.

These results have also been applied to study the low--energy configurations in the Thomson problem corresponding to the local minima obtained 
by perturbing the global minimum, with a subsequent minimization. For the case under study it is found that the global minimum is the configuration, among those studied, which is closer to saturate the bound, which is interpreted as a signal that both strain and energy tend to distribute more uniformly (for the case of the antisymmetric frame potential the lowest energy configuration is not the closest to saturate the bound).

Using the weighted frame potentials, we have also studied the distribution of points in volumes in two and three dimensions, observing that
the bounds are saturated in special cases (for instance hexagonal lattice in two dimension, or cubic lattice in three dimensions).

Among the possible directions of future work we identify:
\begin{itemize}
\item using the spectral decomposition approach to prove further bounds;
\item  perform a systematic study of configurations with a large number of charges based on the bounds obtained;
\end{itemize}

\section*{Acknowledgements}
The  research of P.A. and R.S. is supported by Sistema Nacional de Investigadores (M\'exico).

\begin{appendices}

\section{}
\label{sec:appA}

The non-vanishing coefficients in eq.(\ref{eq_non_uniform_afp})  can be calculated straightforwardly and read:
\begin{equation}
\begin{split}
\kappa_{0,0,0,0,0,0} &= \frac{16 \pi ^{3/2}}{9} \\
\kappa_{2,-1,2,-1,2,2} &= \kappa _{2,-2,2,1,2,1} = \kappa _{2,1,2,-2,2,1} = \kappa _{2,-1,2,2,2,-1} \\
&=  \kappa _{2,2,2,-1,2,-1} = \kappa _{2,1,2,1,2,-2} = \frac{16}{15} \sqrt{\frac{2}{15}} \pi ^{3/2} \\
\kappa _{0,0,2,-1,2,1} &= \kappa _{2,-1,0,0,2,1} =  \kappa _{0,0,2,1,2,-1} =  \kappa _{2,1,0,0,2,-1} \\
&=  \kappa _{2,-1,2,1,0,0} = \kappa _{2,1,2,-1,0,0} = \frac{16 \pi ^{3/2}}{45} \\
\kappa _{2,0,2,0,2,0} &= \frac{32 \pi ^{3/2}}{45 \sqrt{5}} \\
 \kappa _{2,-1,2,0,2,1} &= \kappa _{2,0,2,-1,2,1} = \kappa _{2,-1,2,1,2,0} = \kappa _{2,1,2,-1,2,0} \\ 
&=  \kappa _{2,0,2,1,2,-1} =  \kappa _{2,1,2,0,2,-1} = -\frac{16 \pi ^{3/2}}{45 \sqrt{5}} \\
 \kappa _{2,-2,2,0,2,2} &= \kappa _{2,0,2,-2,2,2} =  \kappa _{2,-2,2,2,2,0} = \kappa _{2,2,2,-2,2,0}  \\
&=  \kappa _{2,0,2,2,2,-2}  = \kappa _{2,2,2,0,2,-2} = -\frac{32 \pi ^{3/2}}{45 \sqrt{5}} \\
 \kappa _{0,0,2,-2,2,2} &=  \kappa _{2,-2,0,0,2,2} = \kappa _{0,0,2,0,2,0} = \kappa _{2,0,0,0,2,0} \\
&= \kappa _{0,0,2,2,2,-2} = \kappa _{2,2,0,0,2,-2} = \kappa _{2,-2,2,2,0,0} \\
&=   \kappa _{2,0,2,0,0,0} = \kappa _{2,2,2,-2,0,0} = -\frac{16 \pi ^{3/2}}{45} \\ 
\end{split} \nonumber
\end{equation}

\end{appendices}

\end{document}